\documentclass[aps,prl,preprint]{revtex4-1}
\usepackage{amsmath}
\usepackage{amssymb}
\usepackage{float}
\usepackage{placeins}
\usepackage{graphicx}
\graphicspath{{./Figures/}}

\begin{document}

\title{Anomalous diffusion and response in branched systems:
        a simple analysis}


\author{Giuseppe Forte}
\address{Dipartimento di Fisica Universit\`a di Roma ``Sapienza'',
P.le A. Moro 2, I-00185 Roma, Italy.}

\author{Raffaella Burioni}
\address{Dipartimento di Fisica and INFN Universit\`a di Parma,
Viale G.P.Usberti 7/A, I-43100 Parma, Italy.}

\author{Fabio Cecconi}
\address{CNR-Istituto dei Sistemi Complessi (ISC),
Via dei Taurini 19, I-00185 Roma, Italy.}

\author{Angelo Vulpiani}
\address{Dipartimento di Fisica Universit\`a di Roma ``Sapienza'' and 
CNR-Istituto dei Sistemi Complessi (ISC), P.le A. Moro 2, I-00185 Roma, 
Italy.}

\date{\today}

\begin{abstract}
We revisit the diffusion properties and the mean drift induced by an
external field of a random walk process in a
class of branched structures, as the comb lattice and the linear
chains of plaquettes.
A simple treatment based on scaling arguments is able to predict the
correct anomalous regime for different topologies.
In addition, we show that even in the presence of anomalous diffusion,
the Einstein's relation still holds, implying a
proportionality between the mean-square displacement of the
unperturbed systems and the drift induced by an external forcing.
\end{abstract}


\maketitle

\section{Introduction \label{sec:Intro}}
The Einstein's work on Brownian motion represents one of brightest 
example of how Statistical Mechanics \cite{Einstein} operates by 
providing the first-principle foundation to phenomenological laws. 
In his paper, the celebrated relationship between the diffusion
coefficient and the Avogadro's number $N_A$ was the 
first theoretical evidence on the validity of the atomistic 
hypothesis. In addition, he derived the first example of a fluctuation 
dissipation relation (FDR) \cite{Kubo,Bettolo08}.

Let $x_t$ be the position of a colloidal particle at time $t$ 
undergoing collisions from small and fast moving solvent particles, 
in the absence of an external forcing. At large times we have:
\begin{equation}
\langle x_t\rangle_0 = 0, \qquad \langle x_t^2\rangle_0 \simeq 2 D\,t \;,
\end{equation}
where $D$ is the diffusion coefficient and the   
average $\langle\cdots\rangle_0$ is over an ensemble 
of independent realizations of the process. 
The presence of an external constant force-field $F$ 
induces a linear drift
\begin{equation}
\langle \delta x_t\rangle_F = 
\langle x_t\rangle_F - \langle x_t\rangle_0 = \mu F t
\end{equation}
where $\langle\cdots\rangle_F$ denotes the average over the perturbed system 
trajectories and $\mu$ indicates the mobility. 
Einstein was able to prove that the following remarkable relation holds: 
\begin{equation}
\frac{\langle x_t^2\rangle_0}
{\langle x_t\rangle_F - \langle x_t\rangle_0} = \frac{2 k_B T}{F}\;.
\label{eq:FDR}
\end{equation}
The above equation is an example of a class of general relations 
known as Fluctuation Dissipation Relations, whose important 
physical meaning is the 
following: the effects of small perturbations on a system can be understood 
from the spontaneous fluctuations of the unperturbed 
system~\cite{Kubo,Bettolo08}.
 
Anomalous diffusion is a well known phenomenon 
ubiquitous in Nature \cite{anomal_Rev90,Castiglione99,Klafter_book}
characterized by an asymptotic mean square displacement behaving as
\begin{equation}
\langle x_t^2\rangle_0 \simeq t^{2\nu} 
\quad\quad 
\mbox{with~} \nu \ne \frac{1}{2}\;.
\label{eq:anomal}
\end{equation} 
The case $\nu> 1/2$ is called superdiffusive, whereas $\nu < 1/2$ 
corresponds to subdiffusive regimes.
The nonlinear behaviour \eqref{eq:anomal} occurs in situations 
whereby the Central Limit Theorem does not apply to the  process 
$x_t$. This happens in the 
presence of strong time correlations and can be found  
in chaotic dynamics~\cite{Geisel84,klages_AnTrBook},
amorphous materials \cite{Amorph} and
porous media \cite{Berkowitz2000,Koch88} as well.

Anomalous diffusion is not an exception also in biological 
contexts, where it can be observed, for instance, in the transport of 
water in organic tissues 
\cite{kopf1996anomalous,Tortuosity2} or migration of 
molecules in cellular cytoplasm \cite{Cytoplasm,Cell_Transp}.   
Biological environments which are crowded with obstacles, compartments 
and binding sites are examples of media strongly deviating from 
the usual Einstein's scenario. 
Similar situations occur when the random walk (RW) is restricted on peculiar topological structures  
\cite{Ben-Avraham,Weiss_Comb,Weiss_Shlomo87}, where subdiffusive behaviours
spontaneously arise.  
In such conditions, it is rather natural 
to wonder whether the fluctuation-response relationship~(\ref{eq:FDR}) 
holds true and, if it fails, what are its possible generalizations. 

\begin{figure}[htbp!]
\centering
\includegraphics[clip=true,keepaspectratio,width=0.3\textwidth,angle=90]
{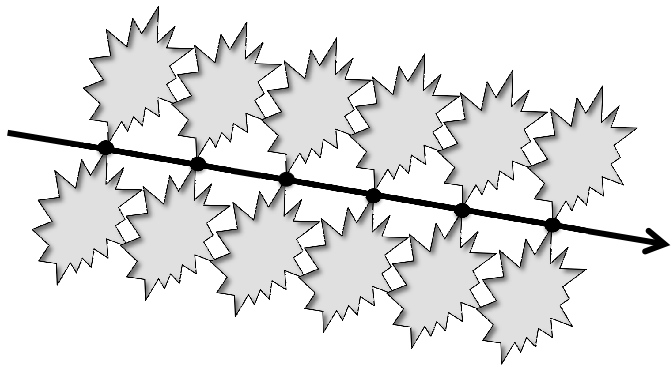}
\caption{Cartoon of a one dimensional lattice 
(backbone) decorated by identical arbitrary-shaped  
sidebranches or dead-ends depicted as lateral irregular objects. 
Such sidebranches act as temporary traps for the random walk along the 
backbone.}
\label{fig:sb}
\end{figure}

The goal of this paper is to present a derivation 
based on a simple physical reasoning, i.e. without 
sophisticated mathematical formalism, of both 
the anomalous exponent $\nu$ and Eq.~(\ref{eq:FDR}) 
for RWs on a class of comb-like and branched structures 
\cite{Burioni05} consisting of a main backbone decorated by an 
array of sidebranches as in Fig.~\ref{fig:sb}. 
Such branched topology is typical of percolation clusters at criticality, 
which can be viewed as finitely ramified fractals 
\cite{coniglio81,coniglio82}. 
Comb-like structures moreover are frequently observed in 
condensed matter and biological frameworks: they describe       
the topology of polymers \cite{polycomb,polybranch}, in particular of
amphiphilic molecules, and can be also engineered at the nano and microscale. 
Moreover, they are studied as a simple models for channels in porous media and 
a general account for these systems can be found in Ref.~\cite{Ben-Avraham}.

The diffusion along the backbone, {\em longitudinal diffusion}, 
can be strongly influenced by the shape and the size of such 
branches and anomalous regimes arise by simply tuning their geometrical 
importance over the backbone. In other words, the  
dangling lateral structures, dead-ends, introduce a delay mechanism 
in the hopping to neighbour backbone-sites that 
easily leads to non Gaussian behaviour, as it was observed for instance in 
flows across porous media \cite{conigliostanley,Hoffmann_RepSiepCarp}.

The simple analysis of the RW on such lattices is 
based on the {\em homogenization time}, meant as the shortest timescale 
after which the longitudinal diffusion becomes standard.
The homogenization time $t_*(L)$ can be identified with the 
typical time taken by the walker to visit most of the 
$M_{sb}(L)$ sites 
in a single sidebranch of linear size $L$. 
Such a time is expected to be a growing function of $M_{sb}(L)$ and thus 
of $L$:   $t_*(L) = g[M_{sb}(L)]$. 
In the following, we shall see how the 
scaling properties of $t_*(L) = g[M_{sb}(L)]$ can be easily extracted 
from graph-theoretical considerations, in simple and complex structures 
as well.

Once such a scaling is known, we can apply a ``matching argument'' 
to derive the exponent $\nu$ in the relation~\eqref{eq:anomal}.
For finite-size sidebranches indeed, the anomalous regime in the 
longitudinal diffusion is transient and soon or later it will be 
replaced by the standard diffusion, 
\begin{equation}
\langle x_t^2 \rangle \sim 
\begin{cases}
t^{2\nu}     & \mbox{if  $t\ll t_*(L)$} \\
D(L)\,t      & \mbox{if  $t\gg t_*(L)$} 
\end{cases}
\label{eq:arg2}
\end{equation}
where $D(L)$ is the effective diffusion coefficients depending 
on $L$.
The power-law and the linear behaviors have to match at time 
$t \sim t_*(L)$, thus we can write the {\em matching condition} 
\begin{equation}
t_*(L)^{2\nu} \sim D(L)\:t_*(L)\quad\mbox{or equivalently}
\quad t_*(L)^{2\nu-1} \sim D(L)\,, 
\label{eq:matching}
\end{equation}  
accordingly, both the scaling  $D(L) \sim L^{-u}$ and $t_*(L) \sim L^v$
provide a direct access to the exponent $\nu$ via the expression 
$(1-2\nu)v = u$.
We shall see in the following, how the values of $u$ and $v$ are determined 
by two relevant dimensions of RW problem: the spectral 
($d_S$) and the fractal ($d$) dimensions. The former is related to return 
probability to a given point of the RW and the latter defines the scaling 
$M_{sb}(L) \sim L^d$.

Moreover, we will show that the anomalous regimes observed in branched
graphs satisfy the FDR~\eqref{eq:FDR} supporting the view that 
FDR has a larger realm of applicability than Gaussian diffusion,  
as already pointed out by other authors
in similar and different contexts \cite{Villamaina011,Barkai98,
Metzler_PhysRep,Chechkin12}. 
In the branched systems considered in this work, 
the generalization of FDR is due to a perfect 
compensation in the anomalous behaviour of the numerator and the denominator of
the ratio~\eqref{eq:FDR}. 

The paper is organized as follows, in sect.2, we discuss the diffusion 
and the response by starting from the simplest branched structure: 
the classical comb-lattice (Fig.~\ref{fig:comb}), i.e. a straight line 
(backbone) intersected by a series of sidebranches. 
The generalization to more sophisticated "branched structures" 
made of complex and fractal sidebranches is reported in sect.3.
Sect.4 contains conclusions, where, possible links 
of the FDRs here derived to other frameworks are briefly discussed.  
  
\section{The simplest branched structure \label{sec:Comb}}
\begin{figure}[htpb!]
\centering
\includegraphics[clip=true,keepaspectratio,width=0.5\textwidth]
{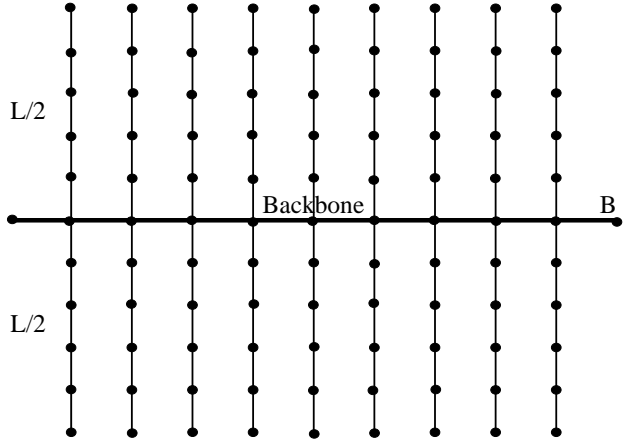}
\caption{Sketch of the simplest comb-lattice structure made of 
a ``backbone'' (horizontal array) and ``tooth'' (lateral arrays) of
size $L$.}
\label{fig:comb}
\end{figure}

At first, we consider the basic model: 
the simplest comb lattice 
is a discrete structure consisting of a periodic and parallel arrangement 
of the ``teeth'' of length $L$ along a ``backbone'' line (B), see 
Fig.~\ref{fig:comb}. 
This model was proposed by Goldhirsch et al. \cite{Goldhirsch_Gefen86} 
as a elementary structure able to describe 
some properties of transport in disordered networks and can be well 
adapted to all physical cases where particles diffuse freely along a 
main direction but can be temporarily trapped by lateral dead-ends.
The walker occupying a site can jump to one of the nearest neighbour sites.
Denoting by $\mathbf r_t = (x_t,y_t)$ the position of the walker at time $t$,
we can define for the longitudinal displacement from the initial position:
\begin{equation}
x_t - x_0 = \sum_{j=1}^t \delta_j
\label{comb:pos_sum}
\end{equation}
where $\{\delta_j\}$ are non independent random variables such that 
$$
\delta_j = 
\left\{
\begin{array}{ll}
\delta^{\parallel}_j  & \mbox{if ~$\mathbf r_j\in$B} \\ 
         0            & \mbox{if ~$\mathbf r_j\notin $B} 
\end{array} 
\right.
$$
where $\delta^{\parallel}_j = \{-1,0,1\}$ with probability $\{1/4,1/2,1/4\}$ 
respectively and $B$ denotes the set of points with $y=0$, i.e. forming the 
backbone B (Fig.~\ref{fig:comb}).  
A simple algebra yields
$$
\langle (x_t - x_0)^2 \rangle_0 = \sum_{j=1}^t \langle \delta_j^2\rangle_0 + 
2\sum_{j=1}^t \sum_{i>j}^t  \langle \delta_j \delta_i \rangle_0
$$
where terms $\langle \delta_j^2\rangle_0 = 0 $ if $\mathbf r_j\notin$B, 
whereas $\langle \delta_j^2\rangle_0 = 1/2$ if $\mathbf r_j\in$B. On the
other hand  $\langle\delta_j\delta_i\rangle_0=0$ for all $j\neq i$.
Therefore we have
\begin{equation}
\langle (x_t - x_0)^2 \rangle_0 = \frac{1}{2} t f_t
\label{eq:comb_msd}
\end{equation}
where $f_t$ is the mean percentage of time (frequency) the walker spends
in the backbone B during the time interval $[0,t]$.
To evaluate $f_t$, we begin from the case $t>t_*(L)$, 
$t_*(L)$  being the {\em homogenization time}, meant as the time taken by the
walker to span a whole tooth, visiting at least once all the sites 
\cite{Weissbook}. 
Since along the $y$-direction the one-dimensional diffusion 
$\langle y_t^2\rangle \simeq 2 D_0 t$ is fast enough to explore 
exhaustively the size $L$ and, more importantly, it 
is recurrent, $t_*(L)$ can be taken as the time such that 
$\langle y_t^2\rangle \sim  L^2$ and thus  $t_*(L) \sim  L^2$.   
Since, after the time of the order $t_*(L) \sim L^2$, 
the probability for the walker to be in a site
of the tooth can be considered to be almost uniform, we have 
$$
f_t = \frac{1}{1+L} \simeq L^{-1}, 
$$
hence for $t\ge t_*(L)$, the mean square displacement behaves as 
\begin{equation}
\langle (x_t - x_0)^2 \rangle_0 \simeq \frac{1}{2(1+L)} t 
\label{eq:comb_stand}
\end{equation}
with an effective diffusion coefficient $D(L) = 1/[4(L+1)]$. In the
above derivation, we have assumed that the lateral teeth are equally 
spaced at distance $1$. When the spacing is $\ell > 1$ the formula changes to 
$D(L) = 1/[4(L+\ell)]$.  
This formula can be interpreted as the ratio between the free $D_0 = 1$ 
and the effective diffusivity $D(L)$. In the literature on transport processes, 
this ratio is sometimes referred to as {\em tortuosity} and
it describes the hindrance posed to the diffusion process by a geometrically
complex medium in comparison to an environment free of obstacles \cite{Tortuosity,Tortuosity2}. 

The diffusion on a simple comb lattice for $L=\infty$ is known 
to be anomalous \cite{Weiss_Comb,anomal_Rev90,Redner}. For finite $L$  
the diffusion remains anomalous as long as the RW does not feel 
the finite size of the sidebranches. 
Therefore for times $t<t_*(L)$, we expect an anomalous 
behaviour 
\begin{equation}
\langle (x_t - x_0)^2 \rangle_0 \sim t^{2\nu}
\label{eq:comb_anomal}
\end{equation}
where the exponent $\nu$ can be computed by the matching 
condition~\eqref{eq:matching}, with $t_*(L) \sim L^2$ and 
$D(L) \sim  L^{-1}$, yielding  $L^{4\nu} \sim L^{-1}\times L^2$, 
from which $\nu = 1/4$, 
\begin{equation}
\langle (x_t - x_0)^2 \rangle_0 \sim t^{1/2}.
\label{eq:comb_exact}
\end{equation} 
This result can be rigorously derived from standard random 
walks techniques~\cite{Weiss_Comb}. 
It is interesting to note that, as the 
homogenization time $t_*(L)$ diverges with the size $L$, upon choosing the 
appropriate $L$, the anomalous regime can be made arbitrarily long till it 
becomes the dominant feature of the process.  

The longitudinal diffusion is a process determined by the return 
statistics of the walkers to the backbone.    
The walker indeed becomes ``active'' only after a return time 
$T_r = T_r(t)$  (operational time) which is actually 
a stochastic variable of the original discrete clock $t=n t_0$. 
This is an example of subordination: the longitudinal diffusion is 
subordinated to a simple discrete-time RW through the operational 
time $T_r$. In a more familiar language, we are observing a       
Continuous Time Random Walk (CTRW) where waiting times are the return 
times to backbone sites \cite{Redner} during the motion along the teeth. 
CTRW on a lattice, proposed by Montroll and Weiss  
\cite{CTRW_latt}, is a generalization of the simple RW 
where jumps among neighbour sites 
do not occur at regular intervals ($t_k = k t_0$) but the 
waiting times between consecutive jumps are distributed according 
to a probability density $\psi(t)$.
Shlesinger \cite{Shlesinger74} showed that anomalous diffusion arises if
$\psi(t)$ is long tailed.

The equation governing the CTRW is
\begin{equation}
P(x,t) = \sum_{n=0}^{\infty} G(x,n) P(n,t)
\label{eq:subo}
\end{equation}
where $G(x,n)$ is the probability distribution of the variable $x$ after 
$n$-steps along the backbone from the origin $x=0$ and $P(n,t)$ indicates 
the probability to 
make exactly $n$-steps in the time interval $[0,t]$. 
The probability $P(n,t)$ is related to the waiting-time distribution $\psi(t)$. 
On the comb lattice, the waiting-time distribution $\psi(t)$ coincides with the 
distribution of first-return time to the backbone sites,  
which for infinite sidebranches is long-tailed and asymptotically decays 
as $\psi(t) \sim t^{-3/2}$ (see \cite{Weiss_Comb}).  
For finite sidebranches of size $L$, the 
distribution is truncated to $t_*(L)$ by the finite-size effect, thus 
$\psi(t) \sim t^{-3/2}\exp[-t/t_*(L)]$, Refs. \cite{anomal_Rev90} and 
\cite{Redner}.

We now consider the problem of the response of a driven RW 
on a comb lattice in the presence of an infinitesimal longitudinal 
(i.e. parallel to the backbone line) external 
field $\epsilon$ \cite{Villamaina011,Giusiano}. 
In that case, the displacement on the backbone is 
$$
x_t - x_0 = \sum_{j=1}^t \Delta_j^{(\epsilon)}
$$
where 
$$
\Delta_j^{(\epsilon)} = 
\left\{
\begin{array}{ll}
\delta_j^{(\epsilon)} & \mbox{if ~$\mathbf r_j\in$B} \\ 
             0        & \mbox{if ~$\mathbf r_j\notin $B} 
\end{array} 
\right.
$$
$\delta_j^{(\epsilon)}=\{-1,0,1\}$ with 
probabilities, $\{(1/4+\delta p),1/2,(1/4-\delta p)\}$, so that  
$\langle \delta_j^{(\epsilon)} \rangle = \epsilon$.
Thus a biased RW with jumping probabilities 
$1/4 - \delta p$ and $1/4 + \delta p$ to the left and 
to the right respectively is used to model the effect of a static 
external field. 
The average jump is   
$\langle \delta_j^{(\epsilon)} \rangle = 1\times (1/4 + \delta p) 
-1\times (1/4 - \delta p) = 2 \delta p$, thus $\epsilon = 2\delta p$.
Notice that $\epsilon$ plays the role of the external field $F$.  
By the same argument used for the free RW on the comb,  
we obtain
\begin{equation}
\langle \delta x_t \rangle_{\epsilon} = 
\langle (x_t - x_0) \rangle_{\epsilon} - \langle (x_t - x_0) \rangle_0 = 
\epsilon t f_t.
\label{eq:comb_drift} 
\end{equation}
The comparison of Eq.~(\ref{eq:comb_msd}) and 
Eq.(\ref{eq:comb_drift}) provides the general result
\begin{equation}
\frac{\langle (x_t - x_0)^2\rangle_0}
{\langle \delta x_t \rangle_{\epsilon}} = 
\frac{1}{2\epsilon}\;. 
\label{eq:comb_response} 
\end{equation}
We stress that this expression holds at any time: 
for both $t\gtrsim t_*(L)$ and $t\lesssim t_*(L)$ \cite{Villamaina011}, thus it works
even when the averages are not taken over the realizations of a
Gaussian process. In this respect, Eq.~\eqref{eq:comb_response}  
represents a generalization of the Einstein's relation 
\eqref{eq:comb_response} to the RW over comb lattices in agreement
with analogous results found in different systems and contexts
\cite{Barkai98,Metzler_PhysRep,Chechkin12}. 

This property is a simple consequence of the subordination condition expressed
by Eq.~\eqref{eq:subo}. 
In fact, the small bias in the left/right jump  ($\epsilon = 2 \delta p$) 
along the  backbone introduces a shift in the distribution of steps 
$$
G_{\epsilon}(x,n) = \frac{1}{\sqrt{2\pi D_s n}} 
\exp \bigg[-\frac{(x - \epsilon n)^2}{2 D_s n} \bigg]
$$ 
where $D_s=1/2$ is the diffusion coefficient of the subordinated 
dynamics $ D_s = \lim_{n\to\infty} 
\langle ({\tilde x}_n - {\tilde x}_0)^2\rangle/(2n)$ and 
${\tilde x}_n$ indicates the position after $n$ jumps on the backbone; 
for $\epsilon = 0$ the distribution is a unbiased Gaussian (in the limit 
of large $t$ also $n$ is large and the Binomial is well approximated by 
Gaussian $G_0(x,n)$).
Actually the precise shape of $G_{\epsilon}(x,n)$ is not very relevant. 
Since $\langle {\tilde x}_n \rangle = \epsilon n$ we can compute the biased 
displacement in the perturbed system
$$
\langle x_t \rangle_{\epsilon} =  
\epsilon \sum_{n=0}^{\infty} P(n,t)\;n\;. 
$$
Considering that $\langle n(t)\rangle = \sum_n P(n,t)\; n$,  
we can re-write
$$
\langle x_t \rangle_{\epsilon} = \epsilon \langle n(t) \rangle\;.
$$
From Eq.~\eqref{eq:subo} we compute the MSD 
obtaining $\langle x_t^2 \rangle_0 = \langle n(t) \rangle/2$   
which is the same result of Eq.~\eqref{eq:comb_msd}, hence
Eq.~\eqref{eq:comb_response} follows.
Note that FDR is exact also for anomalous behaviours as the drift we have 
applied has no effect (or no components) on the sidebranches, therefore 
the waiting time distribution and thus $P_n(t)$ remains unaltered with 
respect to that of the unperturbed system.

To verify the above results, we generated $N_{p}=7\times10^4$ independent 
RW trajectories for $t=2\times 10^7$ time steps over a regular 
comb-lattice with different sidebranch sizes $L$. 
\begin{figure*}[htpb!]
\centering
\includegraphics[clip=true,keepaspectratio,width=0.9\textwidth]
{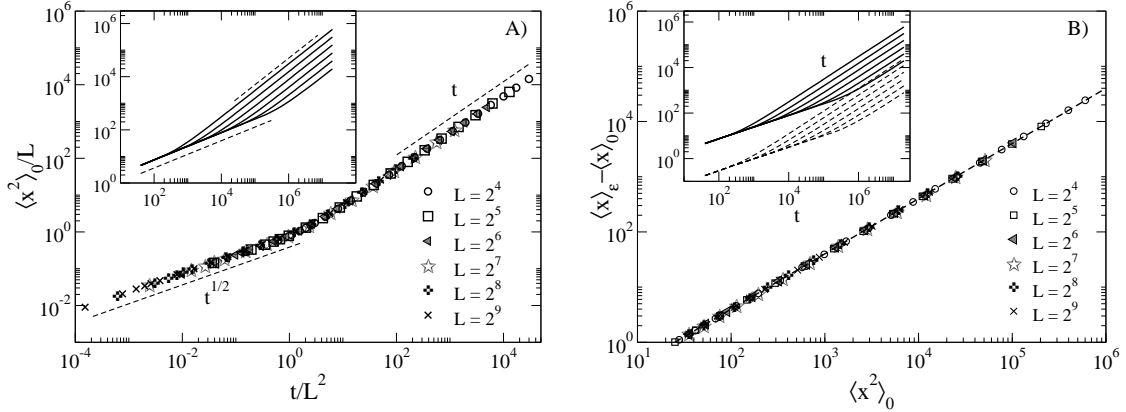}
\caption{\label{fig:1D_Comb}
A) rescaled MSD, $\langle x^2 \rangle_0/L$, 
for the comb lattice (Fig.~\ref{fig:comb}) of tooth length $L$, 
as a function of the rescaled time $t/L^{2}$. 
There is a crossover, at $t/L^{2}\sim 1$ 
(i.e. $t \simeq t_{*} \sim L^{2}$) between a subdiffusive,
$t^{1/2}$, to a standard regime $t$.
{\bf Inset}: plot of $\langle x^2 \rangle_0$ vs. $t$ without rescaling 
for different $L$. 
B) plot showing the generalized fluctuation-dissipation 
relation (\ref{eq:comb_response}). 
The slope of the dashed straight line is $2\epsilon$, 
where $\epsilon=2\delta p = 0.02$ as prescribed by 
Eq.~(\ref{eq:comb_response}). 
{\bf Inset}: separate plot of MSD and fluctuation 
$\langle \delta x_t \rangle_{\epsilon}$ vs. time to appreciate 
their common behaviour in both anomalous and standard regime.}
\end{figure*}
Panel A of Fig.~\ref{fig:1D_Comb} refers to the mean square displacement 
(MSD) for an ensemble of walkers on the traditional comb-lattice  
(Fig.~\ref{fig:comb}) 
at different teeth length to probe the homogenization effects 
characterized by the time $t_{*}(L) \sim L^{2}$.  
The rescaled data ($t/L^{2}$, $\langle x^2 \rangle_0/L$) collapse 
onto a master curve showing a clear crossover, at the rescaled 
crossover time, from a subdiffusive, $t^{1/2}$, to a standard 
regime, $t$. 
The response (panel B of Fig.~\ref{fig:1D_Comb}) 
for the same lattice fulfills the generalized 
fluctuation-dissipation relation~(\ref{eq:comb_response}), 
thus a plot of the response  
$\langle x_t \rangle_{\epsilon} - \langle x_t \rangle_0$ 
vs. the fluctuation $\langle (x_t - x_0)^2\rangle_0$ shows that 
the data for different values of $L$ align along a straight line 
with slope $\epsilon = 2\delta p$. 
The perfect alignment is a consequence of  
the exact compensation at every time between 
fluctuations and response (inset of Fig.~\ref{fig:1D_Comb}B).   
In the simulations of Fig.~\ref{fig:1D_Comb}B, the drift is 
implemented by an unbalance $\delta p=0.01$ in the jump probability 
along the backbone giving $\epsilon = 2\delta p = 0.02$.

\section{Generalized branched structures \label{sec:Gen}}
Interestingly, the previous analysis can be easily extended to the cases 
where each tooth of the comb is replaced by a more complicated 
structure, e.g. a two dimensional plaquette, a cube or a even graph with 
fractal dimension $d$ and spectral dimension $d_{S}$.  
The spectral dimension is defined by the decay of the return 
probability $P(t)$ to a generic site in $t$ steps 
$P(t) \sim t^{-d_S/2}$ \cite{Ben-Avraham,Alexander}, while 
the ratio between $d_S$ and $d$ is known to control the 
mean-square displacement behaviour \cite{Ben-Avraham}
\begin{equation} 
\langle x^{2}(t)\rangle\sim t^{d_S/d}.
\label{eq:msd_graph}
\end{equation}
Of course, formula~(\ref{comb:pos_sum}) still applies to fractal-like 
graphs and Eqs.~(\ref{eq:comb_msd},\ref{eq:comb_response}) hold true, 
provided an 
appropriate change in the ``geometrical'' prefactor is introduced, as we 
explain in the following. 
\begin{figure}[htpb!]
\includegraphics[clip=true,keepaspectratio,width=0.5\textwidth]
{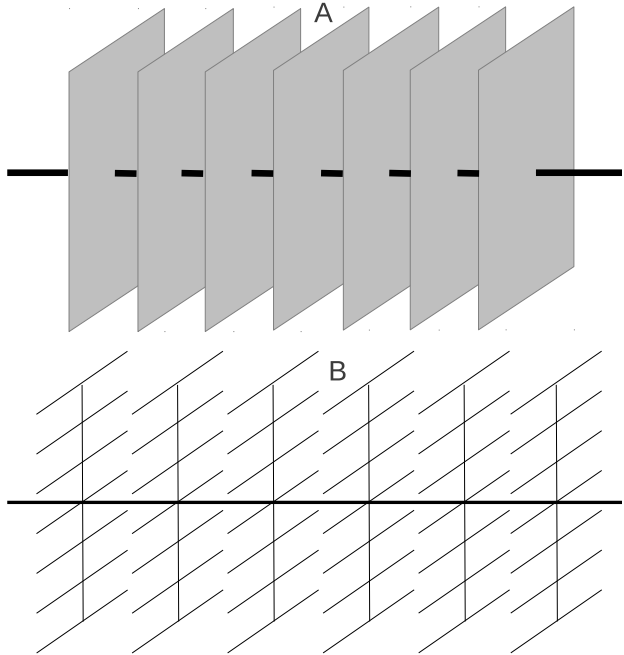}
\caption{Sketch of the comb structures used in the simulations and obtained as 
an infinite periodical arrangement of the same geometrical element:  A) 
comb of plaquettes (dubbed ``kebab'') and B) two-nested comb lattices 
(``antenna'').}
\label{fig:comb_Str}
\end{figure}

In the general case where the ``teeth'' are 
fractal structures with spectral and fractal dimensions  
$d_S$ and $d$ respectively, the lateral diffusion satisfies
$$
\langle y_{t}^{2}\rangle_0 \sim t^{d_{S}/d}.
$$ 
Here, and in the following, $y_t$ indicates the transversal process with 
respect to the backbone.
The previous argument for the homogenization time stems 
straightforwardly by noting that a walker on an infinite sidebranch, 
in an interval $t$, visits a number of different 
sites~\cite{Alexander,Weiss_Comb,Ben-Avraham}
\begin{equation}
M_{sb}(t) \sim
\left\{ 
\begin{array}{ll}
t^{d_S/2}  & \mbox{if  $d_S \leq 2$} \\
t          & \mbox{if  $d_S > 2$}
\end{array}
\right.\, 
\label{eq:Nvisit}
\end{equation}
and accordingly, in a finite sidebranch of linear size $L$, 
the homogenization time $t_*(L)$ is obtained by the condition
$M_{sb}[t_{*}(L)] \sim L^d$ of an almost exhaustive exploration of the sites.
Then when the sidebranch has spectral dimension $d_S \leq 2$, the 
first condition of \eqref{eq:Nvisit} yields  
an homogenization time $t_{*}(L) \sim L^{2 d/d_S}$   
Whereas, if the sidebranch has $d_S>2$, the second condition of 
\eqref{eq:Nvisit} must be
used to obtain $t_{*}(L) \sim L^{d}$.  
The physical reason of a different expression of $t_{*}(L)$ 
above and below $d_S = 2$ is 
due to the non-recurrence of the RW for $d_S>2$ \cite{anomal_Rev90}. 
In this case,   
the  exploration of the sidebranches over a diffusive time scale 
defined by the law \eqref{eq:msd_graph} is not significant 
and the full sampling takes a much longer time which can be estimated 
directly from the second of Eqs.~\eqref{eq:Nvisit}. 

Now using Eq.~\eqref{eq:matching}, we obtain in the case $d_S < 2$
\begin{equation}
\langle (x_{t}-x_{0})^{2}\rangle_0 \sim t^{2\nu},
\qquad
2\nu=1-\frac{d_{S}}{2}.
\label{eq:nu_vs_ds}
\end{equation}
These results coincide with the exact relations obtained by a 
direct calculation of the spectral dimension on branched structures, 
based on the asymptotic behavior of the return probability on the graph, 
or on renormalization techniques \cite{Sofia95,Burioni05,Haynes}.

\begin{figure*}[htpb!]
\centering
\includegraphics[clip=true,keepaspectratio,width=0.8\textwidth]
{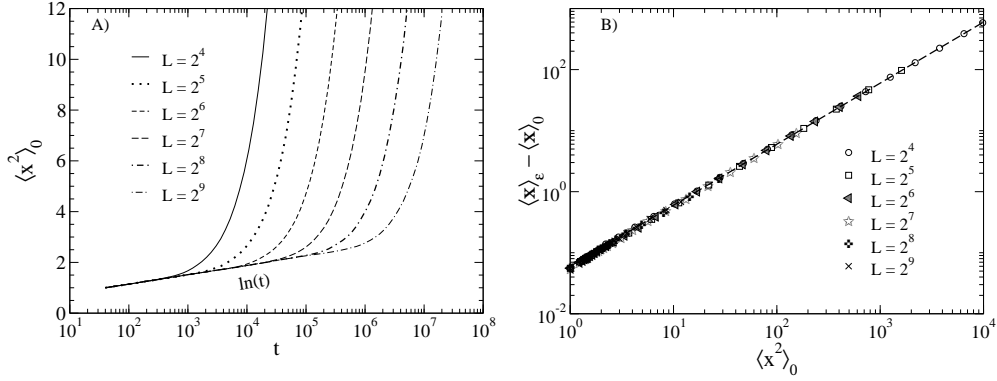}
\caption{Panel A: linear-log plot of MSD, $\langle x_t^2 \rangle_{0}$ vs. time 
for plaquette-comb lattice at different $L$, (Fig.~\ref{fig:comb_Str}A). 
Data show an initial collapse onto 
the common baseline $\ln(t)$, in perfect agreement with the scaling 
result~(\ref{eq:msd_2Dcomb}).
Panel B: plot of the response $\langle \delta x_t \rangle_{\epsilon}$ 
vs. the fluctuations $\langle (x_t-x_0)^2 \rangle_{0}$ showing 
the generalized 
Einstein's relation (\ref{eq:plaquette_response}).
The slope of the dashed line is $3\epsilon$, with $\epsilon = 2\delta p$ 
and $\delta p=0.01$ (unbalance in the left-right jump probability along
the backbone).
\label{fig:2D_Comb}
}
\end{figure*}

The case $d_{S}=2$ deserves a specific treatment thus, as an example, 
we consider the "kebab lattice" 
(Fig. \ref{fig:comb_Str}) where each plaquette is a regular two dimensional 
square lattice, for which  $d_{S}=d=2$. 
Indeed $d_{S}=2$ is the critical dimension separating recurrent ($d_{S} < 2$) 
and not recurrent ($d_{S} > 2$) RWs. Thus $d_{S}=2$ is the marginal 
dimension~\cite{anomal_Rev90} which reflects into the logarithmic scaling
of the transversal MSD $\langle y_{t}^{2}\rangle_0 \sim t/\ln(t)$, hence 
the homogenization time is now $t_{*}(L)\sim L^{2}\ln(L)$. 
Applying once again the matching argument, we obtain the scaling   
\begin{equation}
\langle (x_{t}-x_{0})^{2}\rangle_0 \sim \ln(t)
\label{eq:msd_2Dcomb}
\end{equation}
indicating a
logarithmic pre-asymptotic diffusion along the backbone.
The time evolution of MSD from initial positions
of the simulated random walkers on the ``kebab'' lattice
verifies the transient behaviour \eqref{eq:msd_2Dcomb} 
at different sizes $L$, Fig.~\ref{fig:2D_Comb}A.
 
Notice that, in our matching arguments, we only make use of the spectral 
and fractal dimension of the sidebranches. Interestingly, these two 
parameters are left unchanged if one performs a set of small scale 
transformations on the graph \cite{Univ}, without altering their large scale 
structure. 
Our results hold therefore true also for different and disordered sidebranches, 
provided the two dimensions are unchanged. 
 
Following the same steps as those described for the comb lattice, 
the generalized fluctuation-dissipation relation also holds 
for all branched structures. To check the result we study 
the ``kebab'' lattice (Fig.~\ref{fig:comb_Str}A), 
where the two-dimensional plaquette is a regular square lattice 
of side $L_y$ and unitary spacing \cite{Sofia95}. 
It follows that:
\begin{equation}
\frac{\langle (x_t - x_0)^2\rangle_0}{\langle \delta x_t \rangle_{\epsilon}} = 
\frac{1}{3\epsilon},
\label{eq:plaquette_response} 
\end{equation}
the prefactor $1/3$ stems from the fact that, in a comb-plaquette lattice, 
the probability to jump back and forth along the backbone is $1/6$.
Panel B of Fig.~\ref{fig:2D_Comb} reports the verification of the 
fluctuation-dissipation relation: 
independently of the lattice size, 
the plot 
response vs. MSD is a straight line with slope $1/(3\epsilon)$.

To show the effect of $d_S$ on the homogenization time and on the diffusion 
process, we consider a structure composed by two-nested comb lattices 
that we dub ``antenna'' (Fig.~\ref{fig:comb_Str}B), i.e. a comb lattice, 
where the teeth are comb lattices themselves on the $y,z$ plane. 
This structure is then characterized by two 
length-scales, the vertical, $L_{y}$, and transversal, $L_{z}$, 
teeth length; only for sake of simplicity we 
assume $L_{y} \sim L_{z} \sim L$. 
  
\begin{figure*}[htpb!]
\includegraphics[clip=true,keepaspectratio,width=0.8\textwidth]
{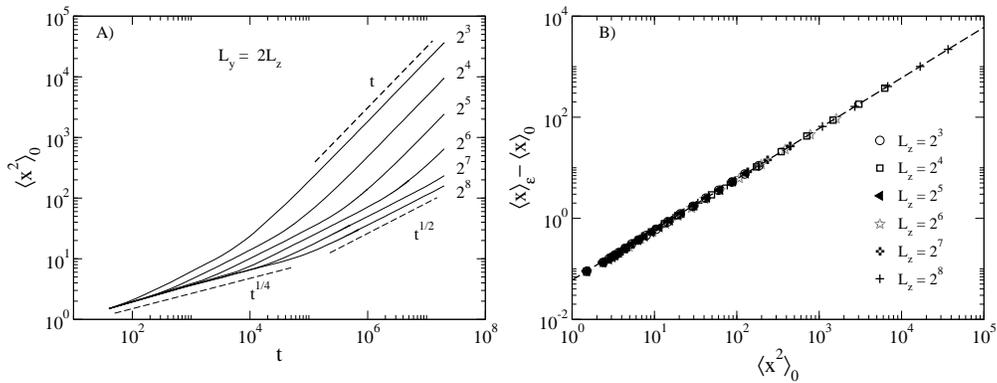} 
\caption{\label{fig:ant_comb} 
A-panel: time behaviour of MSD, $\langle x_{t}^{2} \rangle_0$ 
for the ``antenna'' (Fig.~\ref{fig:comb_Str}B) with 
$L_{y}=2L_{z} = L$ at different values of $L$.
B-panel: generalized fluctuation-dissipation relation 
(\ref{eq:plaquette_response}): response 
$\langle \delta x_{t} \rangle_{\epsilon}$ vs. 
$\langle x_{t}^{2} \rangle_0$. The slope of the dashed straight-line is the 
proportionality factor $3\epsilon$.}
\end{figure*}

Also in this case there exists a crossover time  
$t_{*}(L)\sim L^{2}$ depending on the length of the teeth along-$z$, 
such that:  
for $t\gtrsim t_{*}(L)$, the diffusion becomes 
standard, whereas for $t\lesssim t_{*}(L)$, an anomalous diffusive regime 
takes place.  
Since for a simple comb lattice, $d_S = 3/2$, see \cite{Weiss_Comb},
we obtain from Eq.~(\ref{eq:anomal}) 
$$
\langle (x_{t}-x_{0})^{2}\rangle_0 \sim t^{1/4}\;.
$$
For finite $L$, the MSD in Fig.~\ref{fig:ant_comb}A  
exhibits an initial regime $t^{1/4}$ followed by a $t^{1/2}$-behaviour with 
a final crossover to the standard one. Such a particular scaling, $t^{1/4}$,  
is certainly due to the "double structure" of the sidebranches.  
\FloatBarrier

Also in this case, the generalized Einstein's relation 
is verified (Fig.~\ref{fig:ant_comb}A)
which coincides with Eq.~(\ref{eq:plaquette_response}) for the ``kebab''.
Indeed, the walkers on both antenna and kebab lattices have the 
same probability $1/6$ to make a jump to a nearest neighbour site 
along the backbone. 

\begin{figure*}[t!]
\includegraphics[clip=true,keepaspectratio,width=0.8\textwidth]
{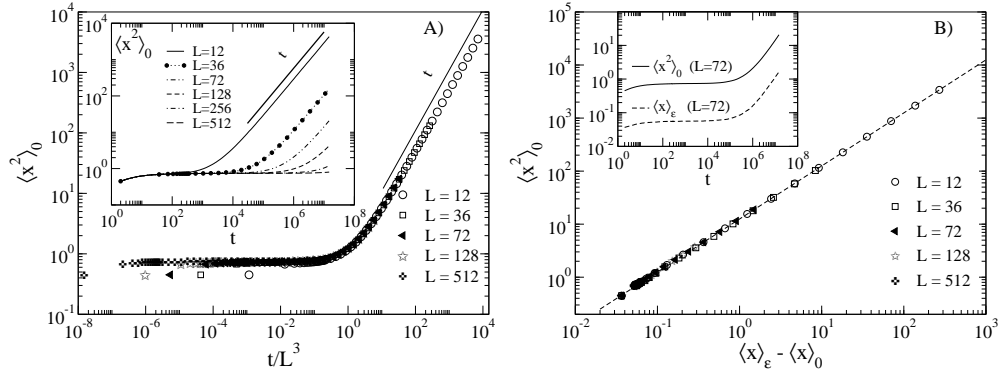}
\caption{\label{fig:cubi_comb} Comb lattice of compenetrating cubes.
A) collapse of the MSD $\langle x_t^2 \rangle_0$ at different cube 
sides $L$ vs. the rescaled time $t/L^{3}$. The data show the plateau 
which is a precursor of the standard diffusion.  
{\bf Inset}: same data not rescaled.
B) Plot showing the generalized fluctuation-dissipation 
relation: response $\langle \delta x_{t} \rangle_{\epsilon}$ vs. 
$\langle x_{t}^{2} \rangle_0$. 
{\bf Inset}: plots of $\langle \delta x_{t} \rangle_{\epsilon}$ and 
$\langle x_{t}^{2} \rangle_0$ showing the parallel behaviour of   
response and MSD independently of the regime.}
\end{figure*}

The case of $d_S>2$ must be carefully 
considered. For simplicity we present our analysis for the particular 
condition $d_S = d = 3$, so we consider a comb-like structure where the 
lateral teeth are compenetrating but non-communicating cubes. For 
computational simplicity the cubes are arranged with centers at a 
unitary distance from one another along the backbone.  
Actually, the minimal distance among the centers of non-compenetrating 
cubes with edge $L$, is $L/2+L/2 = L$ which is of course 
larger than $1$ as soon as $L>1$, but in our model the cubes, despite 
their large overlap, are still considered as distinct sidebranches 
connected only through the backbone.
The homogenization time will be $t_*(L) \sim L^d$ and 
$D(L) \sim L^{-d}$. Therefore, for $t \gg t_*(L)$, we expect the standard 
diffusive growth $\langle (x - x_0)^2 \rangle_0 \sim t/L^{d}$, while 
below $t_*(L)$, $\langle (x - x_0)^2 \rangle_0 \sim t^{2\nu}$ and the 
matching condition at $t_*(L)$ predicts the existence of a plateau 
$\langle (x_t - x_0)^2 \rangle_0 \sim \mbox{const}$, as derived 
by exact relations based on return probabilities \cite{Burioni05}.  
The simulation data are in agreement with the above results,
see Fig.~\ref{fig:cubi_comb}, 
and also the proportionality between fluctuation and response is 
again perfectly verified.


\section{Conclusion \label{sec:Concl}}
In this paper we have analyzed the random walk (RW) and the Einstein's  
response-fluctuation relation 
on a class of branched lattices generalizing the standard comb-lattice. 
For any sidebranch 
of finite-size, a transient regime of anomalous diffusion 
is observed whose exponents can be derived by an heuristic argument
based on the notion of homogenitazion time and on the geometrical 
properties of the lateral structures.
 
Our analysis has been here restricted to branched lattices where the distance
between two consecutive sidebranches is unitary, but it can be
straightforwardly extended to cases with arbitrary spacing.

We can conclude by noting that a random walk 
on generic branched lattice satisfies a generalized Einstein's relation 
for different shapes and sizes $L$ of the 
sidebranches. This is clearly apparent in 
figures:~\ref{fig:2D_Comb}B,~\ref{fig:ant_comb}B and~\ref{fig:cubi_comb}B, 
where data perfectly collapse onto a straight line when plotting 
the free mean square displacements against the response. 

Since this is a straightforward consequence of 
Eqs.~(\ref{eq:comb_msd}) and (\ref{eq:comb_drift}), including their analogues
in more complex comb-structures, the result that  
$$
R(t)  = \frac{\langle (x_t - x_0)^2\rangle_0}
{\langle \delta x_t \rangle_{\epsilon}} = \mbox{const} 
$$
is exact and valid for any comb-like structure both 
in the transient and asymptotic regimes. 
It stems 
from the perfect compensation, at any time, between the response of 
the biased RW and the  mean square displacement of the unbiased RW.

Our results may add other elements to the general issue 
\cite{Debate_Bio,Klages_Book,Ciliberto_FDR} about the validity of the 
fluctuation-dissipation relations (FDR) in far from equilibrium systems 
and non Gaussian transport regimes. 


There are by now sufficient theoretical 
\cite{Barkai98,Villamaina011,Chinappi} and experimental 
\cite{GranuFDR,NGauss_FDR} evidences to claim that 
FDR can be often generalized well beyond its realm of applicability.
This traditional issue of Statistical Mechanics received a renewed 
interest also thanks to the amazing progresses in single-molecule 
manipulation techniques. Experiments whereby a colloidal particle is 
dragged by optical tweezers well approximate the ideal system of 
a single Brownian particle driven out of equilibrium. This   
offers the opportunity to test in a laboratory the FDR on a  
minimal non equilibrium system.  
To some extent, invoking the similarity between RW and Brownian 
motion, the issues addressed in this work involve that class of behaviours
encountered in mesoscopic systems \cite{MolMot}, where either particles or 
generic degrees of freedom move diffusively on a complex support.

\newpage

\bibliographystyle{elsarticle-num}

\bibliography{comb_tras.bib}

\begin{thebibliography}{10}
\expandafter\ifx\csname url\endcsname\relax
  \def\url#1{\texttt{#1}}\fi
\expandafter\ifx\csname urlprefix\endcsname\relax\def\urlprefix{URL }\fi
\expandafter\ifx\csname href\endcsname\relax
  \def\href#1#2{#2} \def\path#1{#1}\fi

\bibitem{Einstein}
A.~Einstein, On the movement of small particles suspended in a stationary
  liquid demanded by the molecular-kinetic theory of heat, Ann. d. Phys. 17
  (1905) 549.

\bibitem{Kubo}
R.~Kubo, The fluctuation-dissipation theorem, Rep. Prog. Phys. 29 (1966) 255.

\bibitem{Bettolo08}
U.~M.~B. Marconi, A.~Puglisi, L.~Rondoni, A.~Vulpiani,
  Fluctuation--dissipation: Response theory in statistical physics, Phys. Rep.
  461 (2008) 111.

\bibitem{anomal_Rev90}
J.~Bouchaud, A.~Georges, Anomalous diffusion in disordered media: statistical
  mechanisms, models and physical applications, Phys. Rep. 195 (1990) 127.

\bibitem{Castiglione99}
P.~Castiglione, A.~Mazzino, P.~Muratore-Ginanneschi, A.~Vulpiani, On strong
  anomalous diffusion, Physica D 134 (1999) 75.

\bibitem{Klafter_book}
J.~Klafter, I.~Sokolov, First Steps in Random Walks, Oxford: Oxford University
  Press, 2011.

\bibitem{Geisel84}
T.~Geisel, S.~Thomae, Anomalous diffusion in intermittent chaotic systems,
  Phys. Rev. Lett. 52 (1984) 1936.

\bibitem{klages_AnTrBook}
R.~Klages, G.~Radons, I.~M. Sokolov, Anomalous transport, Wiley-VCH, 2008.

\bibitem{Amorph}
W.~Schirmacher, M.~Prem, J.-B. Suck, A.~Heidemann, Anomalous diffusion of
  hydrogen in amorphous metals, Europhys. Lett. 13 (1990) 523.

\bibitem{Berkowitz2000}
B.~Berkowitz, H.~Scher, S.~E. Silliman, Anomalous transport in
  laboratory-scale, heterogeneous porous media, Water Resources Research 36
  (2000) 149--158.

\bibitem{Koch88}
D.~L. Koch, J.~F. Brady, Anomalous diffusion in heterogeneous porous media,
  Physics of Fluids 31 (1988) 965.

\bibitem{kopf1996anomalous}
M.~K{\"o}pf, C.~Corinth, O.~Haferkamp, T.~Nonnenmacher, Anomalous diffusion of
  water in biological tissues, Biophys. J. 70 (1996) 2950.

\bibitem{Tortuosity2}
J.~Hrabe, S.~Hrab{\u{e}}tov{\'a}, K.~Segeth, A model of effective diffusion and
  tortuosity in the extracellular space of the brain, Biophys. J. 87 (2004)
  1606--1617.

\bibitem{Cytoplasm}
M.~Weiss, E.~Markus, K.~Fredrik, N.~Tommy, Anomalous subdiffusion is a measure
  for cytoplasmic crowding in living cells, Biophys. J. 87 (2004) 3518.

\bibitem{Cell_Transp}
A.~Caspi, R.~Granek, M.~Elbaum, Diffusion and directed motion in cellular
  transport, Phys. Rev. E 66 (2002) 011916--(12).

\bibitem{Ben-Avraham}
D.~ben Avraham, S.~Havlin, Diffusion and Reactions in Fractals and Disordered
  Systems, Cambridge University Press, Cambridge, 2000.

\bibitem{Weiss_Comb}
G.~H. Weiss, S.~Havlin, Some properties of a random walk on a comb structure,
  Physica A 134 (1986) 474.

\bibitem{Weiss_Shlomo87}
G.~H. Weiss, S.~Havlin, Use of comb-like models to mimic anomalous diffusion on
  fractal structures, Philos. Mag. B 56 (1987) 941--947.

\bibitem{Burioni05}
R.~Burioni, D.~Cassi, Random walks on graphs: ideas, techniques and results, J.
  Phys. A: Math. Gen. 38 (2005) R45.

\bibitem{coniglio81}
A.~Coniglio, Thermal phrase transition of the dilute s-state potts and n-vector
  models at the percolation threshold, Phys. Rev. Lett. 46 (1981) 250.

\bibitem{coniglio82}
A.~Coniglio, Cluster structure near the percolation threshold, J. Phys. A:
  Math. Gen. 15 (1982) 3829.

\bibitem{polycomb}
E.~F. Casassa, G.~C. Berry, Angular distribution of intensity of rayleigh
  scattering from comblike branched molecules, J. Polym. Sci. A-2 Polym. Phys.
  4 (1966) 881--897.

\bibitem{polybranch}
J.~F. Douglas, J.~Roovers, K.~F. Freed, Characterization of branching
  architecture through" universal" ratios of polymer solution properties,
  Macromolecules 23 (1990) 4168--4180.

\bibitem{conigliostanley}
H.~Stanley, A.~Coniglio, Flow in porous media: The backbone fractal at the
  percolation threshold, Phys. Rev. Lett. 29 (1984) 522.

\bibitem{Hoffmann_RepSiepCarp}
S.~Tarafdar, A.~Franz, C.~Schulzky, K.~H. Hoffmann, Modelling porous structures
  by repeated sierpinski carpets, Phys. A: Statistical Mechanics and its
  Applications 292 (2001) 1--8.

\bibitem{Villamaina011}
D.~Villamaina, A.~Sarracino, G.~Gradenigo, A.~Puglisi, A.~Vulpiani, On
  anomalous diffusion and the out of equilibrium response function in
  one-dimensional models, J. Stat. Mech. Theor. Exp. 2011 (2011) L01002.

\bibitem{Barkai98}
E.~Barkai, V.~Fleurov, Generalized einstein relation: A stochastic modeling
  approach, Phys. Rev. E 58 (1998) 1296.

\bibitem{Metzler_PhysRep}
R.~Metzler, J.~Klafter, The random walk's guide to anomalous diffusion: a
  fractional dynamics approach, Phys. Rep. 339 (2000) 1.

\bibitem{Chechkin12}
A.~V. Chechkin, F.~Lenz, R.~Klages, Normal and anomalous fluctuation relations
  for gaussian stochastic dynamics, J. Stat. Mech. Theor. Exp. 2012 (2012)
  L11001.

\bibitem{Goldhirsch_Gefen86}
I.~Goldhirsch, Y.~Gefen, Analytic method for calculating properties of random
  walks on networks, Phys. Rev. A 33 (1986) 2583.

\bibitem{Weissbook}
G.~H. Weiss, Aspects and Applications of the Random Walk, North-Holland,
  Amsterdam, 1994.

\bibitem{Tortuosity}
A.~Koponen, M.~Kataja, J.~Timonen, Tortuous flow in porous media, Phys. Rev. E
  54 (1996) 406--410.

\bibitem{Redner}
S.~Redner, A guide to first-passage processes, Cambridge University Press,
  2001.

\bibitem{CTRW_latt}
E.~W. Montroll, G.~H. Weiss, Random walks on lattices. ii, J. Math. Phys. 6
  (1965) 167.

\bibitem{Shlesinger74}
M.~F. Shlesinger, Asymptotic solutions of continuous-time random walks, J.
  Stat. Phys. 10 (1974) 421--434.

\bibitem{Giusiano}
R.~Burioni, D.~Cassi, G.~Giusiano, S.~Regina, Anomalous diffusion and hall
  effect on comb lattices, Phys. Rev. E 67 (2003) 016116.

\bibitem{Alexander}
S.~Alexander, R.~Orbach, Density of states on fractals: fractons, J. Phys.
  Lett-Paris 43 (1982) 625.

\bibitem{Sofia95}
D.~Cassi, S.~Regina, Random walks on kebab lattices:. logarithmic diffusion on
  ordered structures, Mod. Phys. Lett. B 9 (1995) 601.

\bibitem{Haynes}
C.~Haynes, A.~Roberts, Continuum diffusion on networks: Trees with
  hyperbranched trunks and fractal branches, Phys. Rev. E 79 (2009) 031111.

\bibitem{Univ}
R.~Burioni, D.~Cassi, Geometrical universality in vibrational dynamics, Mod.
  Phys. Lett. B 11 (1997) 1095.

\bibitem{Debate_Bio}
A.~W.~C. Lau, B.~D. Hoffman, A.~Davies, J.~C. Crocker, T.~C. Lubensky,
  Microrheology, stress fluctuations, and active behavior of living cells,
  Phys. Rev. Lett. 91 (2003) 198101.

\bibitem{Klages_Book}
H.~G. Schuster, R.~Klages, W.~Just, C.~Jarzynski, Nonequilibrium Statistical
  Physics of Small Systems: Fluctuation Relations and Beyond, Wiley-VCH, 2013.

\bibitem{Ciliberto_FDR}
S.~Ciliberto, R.~Gomez-Solano, A.~Petrosyan, Fluctuations, linear response, and
  currents in out-of-equilibrium systems, Annu. Rev. of Cond. Matt. Phys. 4
  (2013) 235--261.

\bibitem{Chinappi}
M.~Chinappi, E.~De~Angelis, S.~Melchionna, C.~Casciola, S.~Succi, R.~Piva,
  Molecular dynamics simulation of ratchet motion in an asymmetric nanochannel,
  Phys. Rev. Lett. 97 (2006) 144509.

\bibitem{GranuFDR}
G.~D'Anna, P.~Mayor, A.~Barrat, V.~Loreto, F.~Nori, {Observing brownian motion
  in vibration-fluidized granular matter}, Nature {424} ({2003}) {909--912}.

\bibitem{NGauss_FDR}
Q.~Gu, E.~Schiff, S.~Grebner, F.~Wang, R.~Schwarz, Non-gaussian transport
  measurements and the einstein relation in amorphous silicon, Phys. Rev. Lett.
  76 (1996) 3196--3199.

\bibitem{MolMot}
L.~Le~Goff, F.~Amblard, E.~M. Furst, Motor-driven dynamics in actin-myosin
  networks, Phys. Rev. Lett. 88 (2001) 018101.

\end{thebibliography}

\end{document}